\documentclass{conm-p-l}
\usepackage{amsmath, amssymb, amsthm}

\theoremstyle{plain}
\newtheorem{prop}{Proposition}[section]

\newtheorem{lemm}[prop]{Lemma}
\newtheorem{ques}[prop]{Question}

\theoremstyle{definition}
\newtheorem{defi}[prop]{Definition}

\numberwithin{equation}{section}

\def\Reff#1; #2; #3; #4; #5; #6; #7\par{%
\bibitem{#1} #2, {\it #3}, #4 {\bf #5} (#6) #7}

\def\Ref#1; #2; #3; #4\par{%
\bibitem{#1} #2, {\it #3}, #4}

\newcommand{\Bi}{B_\infty}

\let\ge=\geqslant
\newcommand{\ie}{{\it i.e.}}
\newcommand{\inv}{^{-1}}
\newcommand{\LD}{\mathrm{(LD)}}
\let\le=\leqslant
\newcommand{\op}{\mathbin{*}}
\newcommand{\sh}{\mathrm{d}}
\renewcommand{\ss}[1]{\sigma_{#1}}
\newcommand{\sss}[1]{\sigma_{#1}^{-1}}
\renewcommand{\SS}{S}
\newcommand{\svar}{\widetilde{s}}
\newcommand{\xx}{x}
\newcommand{\yy}{y}

\newcommand{\ZZ}{\mathbb{Z}}
\begin{document}$$$$

\author{Patrick DEHORNOY}
\address{Laboratoire de Math\'ematiques Nicolas
Oresme UMR 6139\\ Universit\'e de Caen,
14032~Caen, France}
\email{dehornoy~math.unicaen.fr}
\urladdr{//www.math.unicaen.fr/$\sim$dehornoy}

\title{Using shifted conjugacy in braid-based cryptography}



\begin{abstract}
Conjugacy is not the only possible primitive for
designing braid-based protocols. To illustrate this
principle, we describe a Fiat--Shamir-style authentication
protocol that be can be implemented using any binary
operation that satisfies the left self-distributive law.
Conjugation is an example of such an operation, but there
are other examples, in particular the shifted conjugation
on Artin's braid group~$\Bi$, and the finite Laver
tables. In both cases, the underlying structures
have a high combinatorial complexity, and they lead to
difficult problems.
\end{abstract}

\maketitle

Most of the braid-based cryptographic schemes proposed so far
\cite{AAG, KoL, CKL} rely on the supposed complexity of
the conjugation operation in Artin's braid groups. In this
note, we would like to stress the fact that conjugation is
by far not the only possible primitive operation for
designing braid-based protocols. 

To illustrate this general idea on a concrete example, we
shall discuss an authentication scheme directly reminiscent
of the Fiat--Shamir scheme, and a variant of some scheme
considered in~\cite{SibT} in the case of braids. We show
that such a scheme can be implemented naturally in every 
algebraic system that involves a binary operation that
satisfies the algebraic law $x(yz) = (xy)(xz)$, called
(left) self-distributivity. Conjugation on any group is an
example of such an operation, but there are other
examples, in particular the operation that we call {\it
shifted conjugation} on Artin's braid group~$\Bi$. There
are reasons to think that sfifted conjugation is (much)
more complicated than standard conjugation, and it could
provide a promising alternative primitive for braid-based
cryptography.

We also mention the Laver tables, which provide other examples
of self-distribu¤-tive operations, this time on a finite underlying
domain of size~$2^n$. Again, these combinatorially very
complex structures could provide a valuable platform.
 
\section{A Fiat--Shamir-like authentication scheme} \label{SFiat}

Here we start with the general principle of the Fiat--Shamir authentication scheme, and show that, under rather natural
hypotheses, it can be implemented in any algebraic system
involving a self-distributive binary operation.

\subsection{The general principle}

Let us start with an arbitrary set~$\SS$, and try to
construct an authentication scheme using the elements
of~$\SS$. To this end, we assume that a function $F_s$ of~$\SS$
into itself is attached to each element of~$\SS$ and that there exist efficiently
sampleable distributions on~Ê$\SS$ such that, provided $s$ and $p$
are chosen according to them, the probability that $s$ 
can be retrieved from the pair $(p, F_s(p))$ in feasible running
time is negligible. Under such
hypotheses, we can use $s$ as a private key, and $(p,
F_s(p))$ as a public key.

A natural idea for designing an authentication scheme is to let
the prover appeal to a second, auxiliary (random) key~$r$,
and use
$F_r(s)$ as a disguised version of~$s$. What we need for a
Fiat--Shamir-like authentication scheme is a commitment of the
verifier guaranteeing that $r$ is fixed, {\it and} an equality
witnessing that $F_r(s)$ is connected in some way to~$s$,
{\it via} the commitment of the prover. As the elements~$p$
and~$F_s(p)$ are public, it is natural to use $F_r(p)$ and/or
$F_r(F_s(p))$ as the commitment(s) of the prover.
Indeed, the assumption that $\xx$ cannot be retrieved
from~$(\yy, F_\xx(\yy))$, which is already needed for~$(p,
F_s(p))$, automatically guarantees that $r$ cannot be
retrieved from the commitments of the prover. 

Then what we need is some equality connecting $x = F_r(p)$,
$y = F_r(F_s(p))$, and $s$---in a way that heavily involves~$s$,
\ie, in such a way that the probability for another $\svar$
to give rise to the same equality is negligible. A simple,
but very particular, solution is to require that $F_s$ and
$F_r$ commute: in this case, the connection between~$x$ and
$y$ is just $y = F_s(x)$. This situation is essentially that
considered in~\cite{Sid, KoL}, and it is not suitable
in the current framework as the verifier would have to know
the secret~$s$. 

A more general and flexible solution is to require that
$F_r(F_s(p))$ be connected to~$F_r(p)$ and~$s$ by some
relation of the form $F_r(F_s(p)) = G_{r, s}(F_r(p))$ for some
new function~$G_{r, s}$. A not so special case is when
$G_{r, s}$ is itself of the form $F_{g(r, s)}$ where $g$ is some
mapping of $\SS \times \SS$ into~$\SS$: considering such a
case is natural, because it avoids introducing a new family of
functions and it enables one to work with the functions
$(F_s)_{s \in \SS}$ solely. For the same reason, it is
natural to consider the case when $g(r, s)$ is defined in
terms of the
$F$-functions, typically $g(r, s) = F_r(s)$. This leads to
requiring that the functions~$F_s$ satisfy the condition
\begin{equation}\label{ELD0}
F_r(F_s(p)) = F_{F_r(s)}(F_r(p)),
\end{equation}
and to use this equality for proving authentication.

\subsection{An authentication scheme}

The previous analysis leads to considering the following
authentication scheme.

We assume that $\SS$ is a set and $(F_s)_{s \in \SS}$ is a family
of functions of~$\SS$ to itself that satisfies
Condition~\eqref{ELD0}.  Then the public keys are a pair~$(p,
p')$ of elements of~$\SS$ satisfying $p' = F_s(p)$, while $s$ is
Alice's private key. The authentication procedure consists in
repeating
$k$~times the following three exchanges:

\begin{quote}
{\parindent=0pt\hrulefill

¥ A chooses $r$ in~$\SS$, and sends the {\it commitments} $x =
F_r(p)$ and $x' = F_r(p')$;

¥ B chooses a random bit~$c$ and sends it to A;

¥ For $c = 0$, A sends $y = r$, and B checks $x = F_y(p)$ and
$x' = F_y(p')$;

¥ For $c = 1$, A sends $y = F_r(s)$, and B checks $x' = F_y(x)$.

\vspace{-1mm}\hrulefill}
\end{quote}

The correctness of the scheme directly follows from
Condition~\eqref{ELD0}. Its security relies on the
following assumptions:

$(i)$ It is impossible to retrieve $s$ from the pair $(p, F_s(p))$,
and, more generally, it is impossible to find~$\svar$ satisfying
$F_s(p) = F_{\svar}(p)$; similar assertions hold for the
pairs $(p, F_r(p))$, $(p', F_r(p'))$, and $(F_r(p), F_r(p'))$; 

$(ii)$ It is impossible to deduce $s$ from $F_r(s)$ when $r$ is
unknown.

\subsection{Self-distributive operations}\label{SLD}

Specifying an $\SS$-indexed family of mappings of a set~$\SS$
into itself amounts to specifying a binary operation on~$\SS$,
namely the operation~$\op$ defined by $\xx \op \yy
=F_\xx(\yy)$. Conversely, $(F_s)_{s \in \SS}$ is the family of
all left translations for~$(\SS, \op)$. Now, in terms of the
operation~$\op$, Condition~\eqref{ELD0} becomes
\begin{equation}\label{ELD}
r \op (s \op p) =(r \op s) \op (r \op p),
\end{equation}
\ie, it asserts that the operation~$\op$ satisfies the {\it left
self-distributivity} law, usually denoted~$\LD$ \cite{Dgd}.

\begin{defi}
A set equipped with a binary operation satisfying \eqref{ELD} is
called an {it LD-system}.
\end{defi}

Translating the previous
authentication scheme  into the language of LD-systems
yields the following version.

Assume that $(\SS, \op)$ is an LD-system.  The public
keys are a pair~$(p, p')$ of elements of~$\SS$ satisfying $p' =
s \op p$, while $s$ is Alice's private key. The authentication
procedure consists in repeating $k$~times the following three
exchanges:

\begin{quote}
{\parindent=0pt\hrulefill

¥ A chooses $r$ in~$\SS$, and sends the commitments $x =
r \op p$ and $x' = r \op p'$;

¥ B chooses a random bit~$c$ and sends it to A;

¥ For $c = 0$, A sends $y = r$, and B checks $x = y \op p$ and
$x' = y \op p'$;

¥ For $c = 1$, A sends $y = r \op s$, and B checks $x' = y
\op x$.

\vspace{-1mm}\hrulefill}
\end{quote}

\section{LD-systems}

The algebraic platforms eligible for implementing the scheme
of Section~\ref{SFiat} are LD-systems, and we are led to
reviewing the existing examples of such algebraic systems.

\subsection{Classical examples}

A trivial example of an LD-system is given by an arbitrary
set~$\SS$ equipped with the operation $x \op y = y$,
or, more generally,
$$x \op y = f(y),$$
where $f$ is any map of~$\SS$ into itself. Such examples are
clearly not relevant for the scheme of Section~\ref{SFiat}, as
the secret~$s$ plays no role in the computation.

The most classical example of an LD-system is provided by
a group~$G$ equipped with the conjugacy operation
$$x \op y = x y x\inv.$$
When $G$ is a non-abelian group for which the conjugacy
problem is sufficiently difficult, $G$ is relevant for the
scheme of Section~\ref{SFiat}, and, more generally, for the
various schemes based on the Conjugacy Search Problem such as
those of~\cite{Sid, KoL} or~\cite{AAG}. Typical platform groups
that have been much discussed in this context are Artin's braid
groups~$B_n$; in particular, the specific scheme considered in
Section~\ref{SFiat} is, in the case of the group~$B_n$,
(a variant of a scheme) proposed by H.\,Sibert in his PhD thesis \cite{SibT}.

\subsection{The shifted conjugacy of braids}

Now, and this is the point we wish to emphasize here,
examples of LD-system of a very different flavour exist. 

Those LD-systems are connected with {\it free} LD-systems,
\ie,  LD-systems that satisfy no other relations than
those resulting from self-distributivity itself. It is
easy to understand that a group equipped with conjugacy,
even a free group, is not a free LD-system: indeed, the
conjugacy operation always satisfies (among others) the
idempotency law $x \op x = x$, and the latter is not a
consequence of~$\LD$, as shows the existence of
non-idempotent LD-system such as the integers equipped with
$x \op y = y + 1$.

Actually, free LD-systems are quite complicated objects, and
we refer to~\cite{Dgd}, which contains an extensive
description. For our purpose, it will be enough to know that, 
for some deep reasons that need not be explained here,
there exists a simple self-distributive operation on
Artin's braid group~$\Bi$ that includes many copies of
the free LD-system with one generator. Let us first recall
the definition \cite{Bir, Dgd}:

\begin{defi} [braid group]
For $n \ge 2$, Artin's {\it braid group} $B_n$ is defined to be
the group with presentation
\begin{equation}\label{EPres}
 \langle \ss1, ..., \ss{n-1} \, ; \,
 \ss i \ss j = \ss j \ss i \; 
 \mbox{for $\vert i - j\vert \ge 2$,}\; 
 \ss i \ss j \ss i = \ss j \ss i \ss j 
 \; \mbox{for $\vert i - j\vert = 1$}
 \rangle.
\end{equation}
\end{defi}

For each~$n$, the identity mapping on~$\{\ss1, ...,
\ss{n-1}\}$ induces an embedding of~$B_n$ into~$B_{n+1}$,
so that the groups~$B_n$ naturally arrange into an inductive
system of groups, and the limit is denoted by~$\Bi$: this is just
the group generated by an infinite family $\ss1, \ss2,...$
subject to the relations~\eqref{EPres}.

\begin{lemm}
Let~$\sh$ be the shift mapping of the sequence $(\ss1,
\ss2, ...)$, \ie, the function mapping~$\ss i$
to~$\ss{i+1}$ for each~$i$. Then $\sh$ induces an
injective morphism of~$\Bi$ into itself.
\end{lemm}

\begin{proof}[Proof (sketch)]
As the relations of~\eqref{EPres} are invariant under shifting
the indices, $\sh$ induces a well-defined endomorphism
of~$\Bi$. That this endomorphism is injective follows
from the interpretation of the elements of~$\Bi$ in terms of
braid diagrams \cite{Bir, Dgd}: the geometric
operation of deleting the leftmost strand is then well-defined, 
and it enables one to deduce $\xx = \yy$ from
$\sh\xx = \sh\yy$.
\end{proof}

The main notion is then the following.

\begin{defi} [shifted conjugacy]
For $\xx, \yy$ in~$\Bi$, we put
\begin{equation}\label{EShift}
\xx \op \yy = \xx \cdot \sh\yy \cdot \ss1 \cdot \sh\xx\inv.
\end{equation}
\end{defi}

The above operation is a skew version of
conjugation: $\yy$ appears in the middle, and it is surrounded
by $\xx$ and $\xx\inv$; the difference with ordinary
conjugation lies in the introduction of the shift~$\sh$,
and of the generator~$\ss1$. The reader can check the
equalities
$$1 \op 1 = \ss1, \qquad
1 \op \ss1 = \ss2\ss1, \qquad
\ss1 \op 1 = \ss1^2 \sss2,\qquad
\ss1 \op \ss1 = \ss2\ss1,$$
which show that shifted conjugation is quite different from
conjugation. 

\begin{prop} \cite{Dfb, Dgd} \label{PDfb}
The system $(\Bi, \op)$ is an LD-system. Moreover, every
braid generates under~$\op$ a free sub-LD-system.
\end{prop}

Checking that the operation defined in~\eqref{EShift} satisfies
the LD law is an easy verification. In the context of
groups, the property that every element generates a free
subgroup is torsion-freeness. Thus Proposition~\ref{PDfb}
expresses that $(\Bi, \op)$ is in some sense a torsion-free
LD-system. 

Understanding why the weird definition of shifted conjugacy
has to appear requires a rather delicate analysis which is
the main subject of the book~\cite{Dgd}. It can be
observed that, once the definition~\eqref{EShift} is used,
braids inevitably appear. Indeed, if we assume that
$G$ is a group, that $f$ is an endomorphism of~$G$, and that
$a$ is a fixed element of~$G$, then defining
$$x \op y = x \, f(y) \, a \, f(x)\inv$$
yields a left self-distributive operation (if and) only if the
subgroup of~$G$ generated by the elements~$f^n(a)$ is a
homomorphic image of Artin's braid group~$\Bi$, \ie, up to an
isomorphism, it is $\Bi$ or a quotient of the latter group.

\subsection{Discussion}

Our intuition is that the LD-system $(\Bi, \op)$, \ie, braids equipped with shifted conjugacy, might be a promising platform for implementing the scheme of Section~\ref{SFiat}---or, more generally, for implementing any scheme based on a left self-distributive operation. This intuition ought to be confirmed by an experimental evidence, which at this early stage is not yet
available. Here we content ourselves with a few remarks about the respective properties of conjugacy and shifted conjugacy in~$\Bi$.

First, note that in general using free structures
does not seem a very good idea in cryptography, as by
definition the free structures are those in which the
least possible number of equalities are satisfied, a not
very good framework for hiding things. That is why, for
instance, a free LD-system would probably not be the
optimal platform for implementing the scheme of
Section~\ref{SFiat}. However, the LD-system~$(\Bi, \op)$
is far from being free, and it is even conjectured that it
contains no free LD-system with two generators. For
instance, the equality $\ss1 \op \ss1 = \ss2 \op \ss2$ ($= \ss2\ss1)$
shows that the sub-LD-system generated by~$\ss1$
and~$\ss2$ is not free. No presentation of~$(\Bi, \op)$ as
an LD-system is known.

Practically, using shifted conjugacy of braids as suggested 
here relies on the difficulty of the following problem,
which is analogous to the Conjugacy Search Problem: 
\begin{quote}
{\bf Shifted Conjugacy Seach Problem}: Assuming that $s, p$ 
are braids in~$\Bi$ and $p' = s \op p$ holds, find a
braid~$\svar$ satisfying $p' = \svar \op p$.
\end{quote}
Contrary to the Conjugacy Seach Problem, no solution to the 
Shifted Conjugacy Search Problem is known so far. It is not
even known whether the simple Shifted Conjugacy Problem is
decidable, \ie, whether one can effectively decide for two
braids~$p, p'$ the existence of~$s$ satisfying $p' = s \op
p$. It is likely that shifted conjugacy is quite different
from ordinary conjugacy, and that none of the many
specific results established for the latter \cite{Gar, FrG,
Geb} extends to shifted conjugacy. In particular, we see
no simple strategy for constructing the ``shifted super
summit set" of a braid~$p$, defined as the family of all
shifted conjugates of~$p$ with minimal canonical
length---which is the key point in all solutions to the
Conjugacy Problem known so far.

However, it is fair to mention that the Shifted Conjugacy 
Search Problem, which should not be threatened by
specific attacks against the Conjugacy Search Problem
\cite{HoS, MSU}, remains, as the latter, an instance of the
general Decomposition Problem and, as such, it is not a
priori immune against length-based attacks \cite{HuT, GaK, GaK2}.

To emphasize the difference between ordinary and shifted 
conjugations, we point

\begin{prop} [\cite{Dgb}, Corollary~1.8]
The mapping $f: s \mapsto s \op 1$ is injective.
\end{prop}

In the case of ordinary conjugacy, every conjugate of~$1$ is~$1$, so the above injective function~$f$ is replaced with the constant function with value~$1$. By the way, very little is known about~$f$. In particular, we raise

\begin{ques}
Starting with a braid~$p$, find~$s$ satisfying $s \op 1 = p$ (when it exists).
\end{ques}

Once more, nothing is known. This might suggest to
use~$f$ as a possible one-way function on braids.

\section{The Laver tables and other algebraic systems}

To conclude, we mention that braids are not the only
possible platform for implementing self-distributive
operations---and that the self-distributivity law is not
even the only algebraic law eligible for the approach
sketched in Section~\ref{SFiat}.

\subsection{The Laver tables}

Instead of resorting to an infinite LD-system like $\Bi$
equip\-ped with shifted conjugacy, one could instead use finite
LD-systems. Such algebraic systems are far from being
completely understood, but there exists an infinite sequence
of so-called Laver tables that plays a fundamental
role among LD-systems---similar to the role of the
cyclic groups~$\ZZ/\!p\ZZ$ among finite abelian groups---and,
at the same time, has a high combinatorial complexity.

We refer to Chapter~X of~\cite{Dgd} for details. For our
current overview, it is enough to mention that, for each
nonnegative integer~$n$, there exists a unique
LD-system~$A_n$ such that the underlying set is the
$2^n$ elements interval $\{0, 1, ..., 2^n-1\}$ and one has $p \op
0 = p+1$ for $0 \le p \le 2^n-2$ and $2^n-1 \op 0 = 0$. The value of $p \op q$ in~$A_n$ can be easily computed using a double
induction on $q$ increasing from~$0$ to~$2^n-1$ and for $p$
decreasing from~$2^n-1$ to~$0$, using the rule 
$$p \op (q + 1) = (p \op q) \op (p \op 0),$$
and observing that $p \op q$ has to be always strictly larger
than~$p$. Table~\ref{TLaver} displays the first four Laver
tables.

\begin{table}[ht]
\begin{tabular}{c|c}
$A_0$&$0$\\
\hline
$0$&$0$
\end{tabular}
\quad
\begin{tabular}{c|cc}
$A_1$&$0$&$1$\\
\hline
$0$&$1$&$1$\\
$1$&$0$&$1$
\end{tabular}
\quad
\begin{tabular}{c|cccc}
$A_2$&$0$&$1$&$2$&$3$\\
\hline
$0$&$1$&$3$&$1$&$3$\\
$1$&$2$&$3$&$2$&$3$\\
$2$&$3$&$3$&$3$&$3$\\
$3$&$0$&$1$&$2$&$3$
\end{tabular}
\quad
\begin{tabular}{c|cccccccc}
$A_3$&$0$&$1$&$2$&$3$&$4$&$5$&$6$&$7$\\
\hline
$0$&$1$&$3$&$5$&$7$&$1$&$3$&$5$&$7$\\
$1$&$2$&$3$&$6$&$7$&$2$&$3$&$6$&$7$\\
$2$&$3$&$7$&$3$&$7$&$3$&$7$&$3$&$7$\\
$3$&$4$&$5$&$6$&$7$&$4$&$5$&$6$&$7$\\
$4$&$5$&$7$&$5$&$7$&$5$&$7$&$5$&$7$\\
$5$&$6$&$7$&$6$&$7$&$6$&$7$&$6$&$7$\\
$6$&$7$&$7$&$7$&$7$&$7$&$7$&$7$&$7$\\
$7$&$0$&$1$&$2$&$3$&$4$&$5$&$6$&$7$
\end{tabular}
\medskip
\caption{\smaller\sf The Laver tables $A_n$ with $0 \le n \le
3$}
\label{TLaver}
\end{table}

Several general phenomena can be observed on these
particular examples. First, for each~$n$, the table~$A_n$
with $2^n$~elements is the projection modulo~$2^n$ of the
table~$A_{n+1}$ with $2^{n+1}$ elements. In other words, if we
use a length~$n$ binary representation for the elements
of~$A_n$, only the dominant bit of each value has to be
computed in order to determine~$A_{n+1}$ from~$A_n$.
Next, every row in the table~$A_n$ is periodic, with a period
that is a power of~$2$. More precisely, for each~$p$, the row
of~$p$ in~$A_n$ consists of $2^k$ values 
$$r_0 = p + 1 < r_1 < ... < r_{2^k-1} = 2^n-1$$
repeated $2^{n-k}$~times. One can show that, if $(r_0,
..., r_{2^k-1})$ is the periodic pattern in the row of~$p$
in~$A_n$, with $r_0 = p+1$ and $r_{2^k-1} = 2^n - 1$ and if
$t$ denotes the smallest integer for which one has $p \op t
\ge 2^n$ in~$A_{n+1}$, then 

- $(i)$ either $t = 2^k$ holds, the period
of~$p$ doubles from~$2^k$ to~$2^{k+1}$ between~$A_n$
and~$A_{n+1}$, and the periodic pattern in~$A_{n+1}$
is $(r_0, ..., r_{2^k-1}, r_0 + 2^n, ..., r_{2^k-1} + 2^n)$,

- $(ii)$ or $0 \le t < 2^k$ holds, the period
of~$p$ remains~$2^k$ in~$A_{n+1}$, and the periodic
pattern in~$A_{n+1}$ is
$(r_0$, ..., $r_{t-1}, r_t + 2^n$, ..., $r_{2^k-1} +
2^n)$.\\ 
In each case, the only piece of information needed
to  construct the row of~$p$ in~$A_{n+1}$ from the row
of~$p$ in~$A_n$ is the value of~$t$, which is called the
{\it threshold} of~$p$ in~$A_n$, and, therefore, the list
of thresholds suffices to construct~$A_{n+1}$
from~$A_n$ ({\it cf.} Table~\ref{TLaver}). Note that, as
$A_n$ is the projection of~$A_{n+1}$, we can consider that
we work in the inverse limit~$A_\infty$ of the
tables~$A_n$, \ie, we are constructing an LD-operation on
$2$-adic numbers.

\begin{table}[htb]
\begin{tabular}{c|c}
$A_0$&$1$\\
\hline
&$-$
\end{tabular}
\quad
\begin{tabular}{c|cc}
$A_1$&$1$&$2$\\
\hline
&$0$&$1$
\end{tabular}
\quad
\begin{tabular}{c|cccc}
$A_2$&$1$&$2$&$3$&$4$\\
\hline
&$1$&$0$&$0$&$2$
\end{tabular}
\quad
\begin{tabular}{c|cccccccc}
$A_3$&$1$&$2$&$3$&$4$&$5$&$6$&$7$&$8$\\
\hline
&$2$&$2$&$1$&$0$&$0$&$0$&$0$&$4$
\end{tabular}
\medskip
\caption{\smaller\sf Threshold table for $A_n$ with $1 \le n \le
3$}
\label{TLaver}
\end{table}

The reason for mentioning the Laver tables here is that their 
combinatorial properties seem to be very complicated. In
particular, predicting the values in the first half of the
sequence of thresholds is extremely difficult (the values
in the second half are always $0$, ..., $0$, $2^n$): this
is witnessed by the results of~\cite{Dou, DoJ, Dra}
which show that fast growing functions are necessarily
involved here.

\subsection{Central duplication}

As a final remark, we come back to the Fiat--Shamir-like
authentication scheme of Section~\ref{SFiat}. We noted that
its security requires two conditions, namely one that is
directly connected with the difficulty of what can be
called the $\op$-Search Problem, and the additional
requirement that communicating $F_r(s)$, \ie, $r \op s$,
gives no practical information about~$s$ when $r$ remains
unknown. Using the latter condition to forge an attack
seems unclear, but, at least for aesthetic reasons, we
might like to avoid it. This can be done, at the expense
of changing the algebraic law.

Indeed, instead of communicating~$F_r(s)$ in case $c=1$ of
the authentication scheme, Alice could communicate $F_s(r)$. 
In this case, the supposed difficulty of the $\op$-Search
Problem guarantees that $F_s(r)$ gives no information
about~$s$. Now, when the scheme is modified in this way,
the equality checked by the verifier has to be modified as
well. If we keep the same principle, we are led to
replace Condition~\eqref{ELD0} with
\begin{equation}\label{ECD0}
F_r(F_s(p)) = F_{F_s(r)}(F_r(p)).
\end{equation}
When Condition~\eqref{ECD0} is translated into the
language of binary operations, we obtain a new algebraic
law, namely
\begin{equation}\label{ECD}
r \op (s \op p) = (s \op r) \op (r \op p),
\end{equation}
instead of left self-distributivity. Nothing specific is known
about this law so far, but it should be possible to use the 
general method explained for a similar law in \cite{Dgj} 
to construct concrete examples of algebras that satisfy it.

\section{Conclusion}

We discussed various non-classical algebraic operations that
could possibly be used as cryptographical primitives, typically
for a Fiat--Shamir-like authentication scheme. The most
promising example seems to be the shifted conjugacy operation
on braids. At the least, the existence of such an operation shows
that conjugacy is not the only possible primitive for 
braid-based cryptography, and that further investigation in this
direction is needed.

\end{document}